\documentclass[conference]{IEEEtran}
\IEEEoverridecommandlockouts

\usepackage{cite}
\usepackage{amsmath,amssymb,amsfonts}
\usepackage{graphicx}
\usepackage{textcomp}
\usepackage{xcolor}
\usepackage{booktabs} 
\usepackage{hyperref} 
\usepackage{subcaption} 
\usepackage{algorithm} 
\usepackage{algorithmic} 
\usepackage{multicol}
\usepackage{acronym}
\newacro{FM}{ferromagnet}
\newacro{HM}{heavy-metal}
\newacro{DMI}{Dzyaloshinskii-Moriya interaction}
\newacro{CIP}{current-in-plane}
\newacro{CPP}{current-perpendicular-to-plane}
\newacro{PMA}{perpendicular magnetic anisotropy}
\newacro{AFM}{anti-ferromagnetic}
\newacro{LIF}{leaky-integrate-and-fire}
\newacro{LTD}{long-term depression}
\newacro{LTP}{long-term potentiation}
\newacro{SNN}{spiking neural network}
\newacro{TL}{top layer}
\newacro{BL}{bottom layer}
\newacro{OOMMF}{Object-Oriented MicroMagnetic Framework}
\newacro{MTJ}{magnetic tunnel junction}
\newacro{DW}{domain wall}
\newacro{FL}{free layer}
\newacro{PL}{pinned layer}
\newacro{LLGS}{Landau-Lifshitz-Gilbert-Slonczewski}
\newacro{STT}{spin-transfer torque}
\newacro{SOT}{spin-orbit torque}
\newacro{SHE}{spin-Hall effect}
\newacro{eNVM}{emerging non-volatile memory}
\newacro{ReRAM}{resistive RAM}
\newacro{MRAM}{magnetoresistive random-access memory}
\newacro{TMR}{tunnel magnetoresistance}
\newacro{WWL}{write word-line}
\newacro{WBL}{write bit-line}
\newacro{SL}{select line}
\newacro{RBL}{read bit-line}
\newacro{MSB}{most significant bit}
\newacro{LSB}{least significant bit}
\newacro{P}{parallel}
\newacro{AP}{anti-parallel}

\usepackage{hyperref}

\usepackage[margin=0.74in]{geometry}

\def\BibTeX{{\rm B\kern-.05em{\sc i\kern-.025em b}\kern-.08em
		T\kern-.1667em\lower.7ex\hbox{E}\kern-.125emX}}

\begin{document}
	
	\title{Edge Detection Framework Utilizing SOT-MTJ Bit-Cell Arrays
		\thanks{This work was supported by the New Faculty Seed Grant(Grant Reference: N4/24/1005) by BITS Pilani K K Birla Goa Campus.}
	}
	
	\author{\IEEEauthorblockN{Kushagra Singh, Debasis Das}
		\IEEEauthorblockA{\textit{Dept. of Electrical and Electronics Engineering} \\
			\textit{BITS Pilani, K K Birla Goa Campus}\\
			Goa, India, 403726 \\
			Email: {f20231210,debasisdas}@goa.bits-pilani.ac.in}
	}
	
	\maketitle
	
	\begin{abstract}
		Traditional edge detection algorithms, foundational to computer vision, face significant challenges in energy efficiency and processing latency on conventional CMOS-based hardware.
		Existing algorithms, such as Canny, are computationally expensive, posing challenges in resource-constrained hardware where energy efficiency and low latency are critical.
		This study introduces a novel, hardware-efficient algorithm that leverages the intrinsic characteristics of \ac{MTJ} devices.
		We present a detailed device-level analysis of an \ac{MTJ}-based system for edge detection, outlining its operational cycles, including write, read, and reset methods.
		The algorithm's efficacy is evaluated against the standard Canny edge detection method.
		We provide a quantitative performance analysis, including metrics such as energy consumption and latency, which demonstrates that our proposed spintronics-based approach offers a promising solution for achieving low-power, high-speed image processing.
	\end{abstract}
	
	
	\section{Introduction}\label{Sec:Intro}
	Edge detection is a cornerstone of digital image processing and computer vision, serving as the initial step for higher-level tasks such as object recognition, image segmentation, and scene understanding~\cite{canny1986computational,maini2009study, vincent2009descriptive,ye2024diffusionedge}.
	Mainstream algorithms, such as the Canny edge detector, have proven effective but are inherently computationally intensive~\cite{maini2009study}.
	This computational burden, however, is particularly problematic for resource-constrained edge devices, where the demand for low power consumption and minimal processing latency is paramount.
	This is largely due to the Von-Neumann bottleneck, which forces a constant, time-consuming data transfer between the separate processing unit and memory, limiting the parallelism and efficiency necessary for data-intensive tasks like image processing.
	As CMOS scaling approaches its physical limits, these issues are exacerbated, motivating the exploration of alternative computing paradigms~\cite{basu2018nonsilicon}.
	To overcome this fundamental architectural limitation, \ac{eNVM} devices-such as \ac{ReRAM}~\cite{zhao2019physics}, ferroelectric memories~\cite{chen2025low}, and spintronic~\cite{chen2023spintronic} devices are being explored. 
    Apart from storing the memory bits, these devices offer in-memory computing, where computational tasks are performed directly within the memory array itself~\cite{chen2016review,roy2020memory,yu2020compute}.
	This approach drastically reduces data movement, allowing for massive parallelism, which is essential for accelerating data-intensive tasks while simultaneously achieving significantly lower power consumption and higher processing speeds.
	Among various \ac{eNVM} devices, spintronics devices are particularly promising as they offer high speed, excellent endurance, low static power consumption, and compatibility with existing CMOS technology~\cite{ralph2008spin}.
	The continued demand for high-speed, low-power, and highly scalable non-volatile memories has driven intense research into \ac{MRAM}.
	The fundamental building block of \ac{MRAM} is the \ac{MTJ}, where a thin insulating barrier is sandwiched between two ferromagnetic layers~\cite{zhu2006magnetic,ikeda2007magnetic}.
	Data is stored by controlling the relative magnetic alignment of these layers.
	
	In this work, we investigate the device physics of \acp{MTJ} to realize a novel hardware-friendly image edge detection algorithm.
	Our primary objective is to exploit the switching characteristics of the \acp{MTJ}, to achieve a low-latency and energy-efficient edge detector.
	We perform numerical simulations of the MTJ's magnetization dynamics to demonstrate its utility as a reconfigurable kernel for image processing.
	The approach is validated by comparing its performance against the industry-standard Canny edge detection algorithm, thereby demonstrating its potential as a next-generation solution for efficient in-sensor or in-memory image processing.
	
	\section{Device Physics and Operation}\label{Sec:Device}
	The core of our device is the \ac{MTJ}, a three-layer stack consisting of an oxide layer, sandwiched between two \ac{FM} layers.
	The first \ac{FM} layer, referred to as the \ac{PL}, has a magnetization direction fixed by an exchange bias to an adjacent antiferromagnetic layer.
	In contrast, the second \ac{FM} layer, the \ac{FL}, is designed with uniaxial anisotropy, allowing its magnetization to be switched between two stable states by an external stimulus.
	The device exhibits two distinct resistance states: a low-resistance state ($\mathrm{R_{P}}$) when the magnetizations of the FL and PL are aligned in \ac{P}, and a high-resistance state ($\mathrm{R_{AP}}$) when they are in \ac{AP} orientation~\cite{moodera1995large}.
	This difference in resistances allows the \ac{MTJ} to function as a non-volatile memory element by associating each resistance state with a binary value (e.g., ``0'' or ``1'').
	The quality of the device is quantified by the \ac{TMR} ratio, which is given as TMR=$\mathrm{\left(R_{AP}-R_P\right)/R_P}$.
	The magnetization of the free layer is altered by passing a spin-polarized current through the device.
	There are two basic methods to perform this operation.
	First, there is the \ac{STT}, where the spin angular momentum from the pinned layer is transferred to the free layer via spin-polarized electrons, causing its magnetization to switch.
	However, a key disadvantage of \ac{STT} is that the switching current must pass directly through the tunnel barrier, which can degrade the barrier over time and reduce the device's endurance.
	Furthermore, the shared read and write current path in \ac{STT} can lead to unwanted bit flips during read operations, compromising data integrity.
	This challenge led to the development of \acf{SOT}, where the spin current is generated in a separate heavy metal layer grown next to the \ac{FL}, and the spin torque is applied in a plane perpendicular to the current~\cite{liu2012spin}.
	This method is advantageous because it separates the read and write current paths, preventing degradation of the tunnel barrier and leading to faster, more energy-efficient switching.	
	
	\begin{figure}[!b]
		\centering
		\includegraphics[scale=0.28]{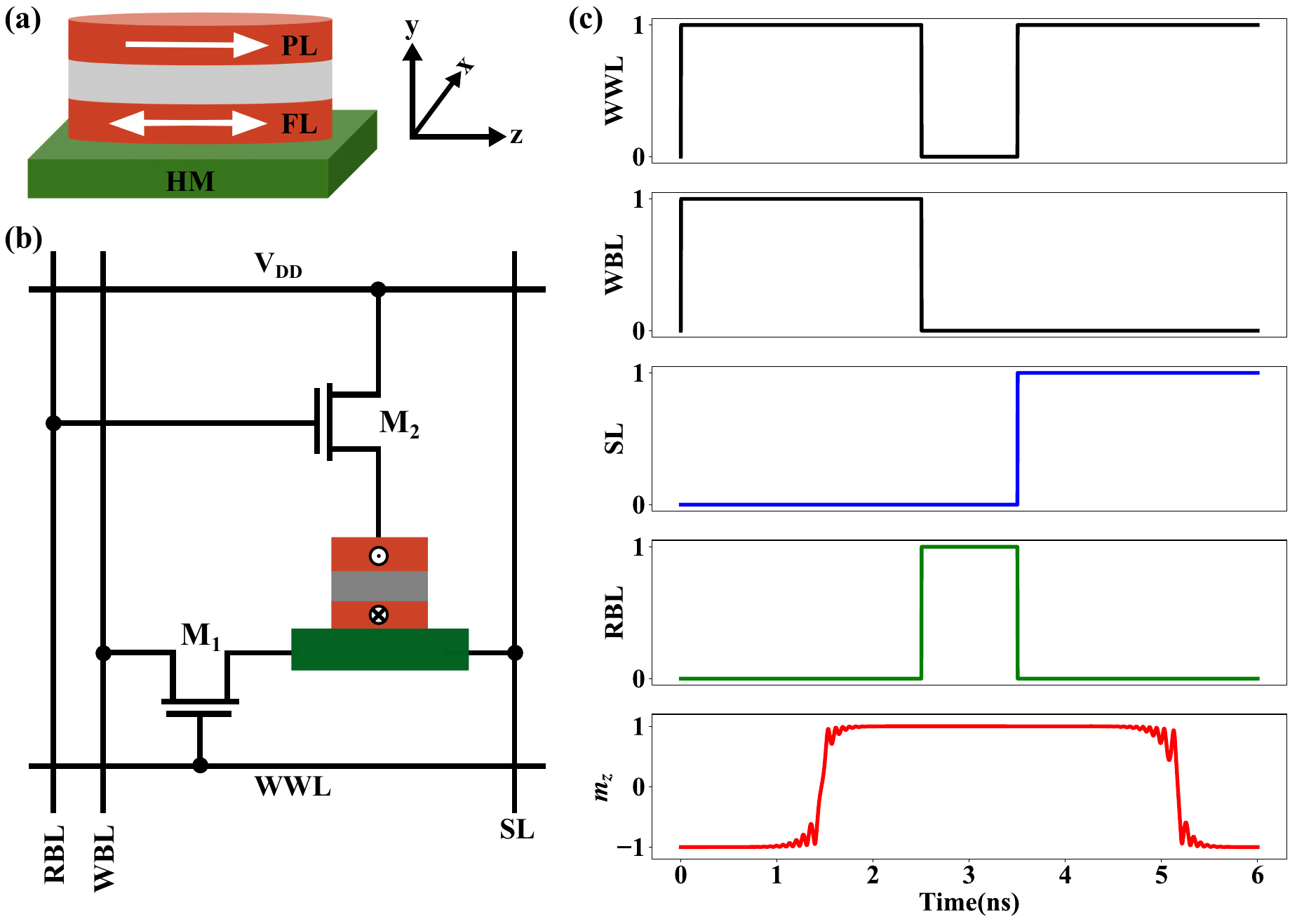}
		\caption{(a) Schematic of the SOT-MTJ device, on top of the \ac{HM} layer. (b) Schematic of the bit-cell of SOT-MTJ with the access transistors for the write, read, and reset mechanisms. (c) Variation of different signals across the bit-cell and the $m_z$ with time.}
		\label{fig:read_write_reset}
	\end{figure}
	
	Fig~\ref{fig:read_write_reset}(a) illustrates the schematic of an SOT-MTJ device, where the \ac{FL} is attached to a \ac{HM} layer.
	Magnetization of the \ac{PL} layer is fixed along the $+z\mbox{-}$direction, while the magnetization of the \ac{FL} is switched by \ac{SOT}. 
	When an electric current is passed through the \ac{HM} layer along the $+x\mbox{-}$direction, the electron spins get polarized along the $\pm z\mbox{-}$direction, where the $+z\mbox{-}$direction ($\mbox{-}z\mbox{-}$direction) spins may accumulate on the top (bottom) surface of the \ac{HM} layer.
	This leads to a spin current flowing into the \ac{FL}, and eventually switching the magnetization.
	Magnetization dynamics of the \ac{FL} are computed by solving the \acf{LLGS} equation numerically, which is given as~\cite{ralph2008spin}
	\begin{equation}
		\frac{d\hat{m}}{dt}=-\gamma \mu_0 \left(\hat{m}\times \mathbf{H}_{eff}\right) +\alpha\left(\hat{m}\times\frac{d\hat{m}}{dt}\right)+\boldsymbol{\tau}_{SOT}.
	\end{equation}
	Here, $\hat{m}$ is the unit vector of the magnetization, $\mathbf{H}_{eff}$ is the effective magnetic field, and $\boldsymbol{\tau}_{SOT}$ is the spin-orbit torque.
	$\mathbf{H}_{eff}$ consists of the uniaxial field, $\mathbf{H}_u=\left(2K_u/\mu_0 M_s\right)m_z \hat{z}$, and the demagnetization field $\mathbf{H}_{demag}=\left(\mathbf{N}\cdot\hat{m}\right)\hat{m}$.
	Spin-orbit torque $\boldsymbol{\tau}_{SOT}$ is given by $-\frac{\gamma\hbar I_s}{2q M_s V}\left(\hat{m}\times\hat{m}\times\boldsymbol{\sigma}\right)$, where $I_s$ is the spin-polarized current, given by $I_s=P_s I_c$, $V$ is the volume of the \ac{FL}.
	$P_s$ is the spin polarization given by the following expression,
	\begin{equation}
		P_s = \left(\frac{A_{MTJ}}{A_{HM}}\right)\theta_{SH}\left(1-sech\left(\frac{t_{HM}}{\lambda_{sf}}\right)\right).
	\end{equation}
	Here, $A_{MTJ}$ is the cross-sectional area of the \ac{MTJ}, across the $x\mbox{-}z$ plane (Fig.\ref{fig:read_write_reset}(a)), and the $A_{HM}$ is the cross-sectional area of the \ac{HM} layer through which the charge current is passed, \emph{i.e.} the $y\mbox{-}z$ plane.
	The rest of the parameter details are given in Table I.

	In Fig.~\ref{fig:read_write_reset}(b), we demonstrate a bit-cell configuration for a single SOT-MTJ device to perform \textit{write}, \textit{read}, and \textit{reset} operations, with the help of two access transistors $\mathrm{M_1}$ and $\mathrm{M_2}$.
	We initially assume the magnetization of the \ac{FL} is oriented along the $\mbox{-}z\mbox{-}$direction.
	The \textit{write} operation is initiated by applying a logic `1' to both the \ac{WWL} and \ac{WBL}, and a logic `0' to the \ac{SL} for a duration of 2.5 ns.
	This voltage configuration activates transistor $\mathrm{M_1}$, which in turn facilitates the flow of current through the \ac{HM} layer.
	Following this, logic `0' is applied to both the \ac{WWL} and \ac{WBL} while the \ac{SL} is held at logic `0', thereby deactivating transistor $\mathrm{M_1}$.
	Subsequently, the read operation is initiated by applying a logic `1' to the \ac{RBL} for 1 ns to activate transistor $\mathrm{M_2}$, which permits a current to flow through the \ac{MTJ}.
	The magnitude of this current is determined by the resistance state of the \ac{MTJ} ($\mathrm{R_{P}}$ or $\mathrm{R_{AP}}$).
	Next, the \ac{RBL} and \ac{WBL} are connected to logic `0', while the \ac{SL} and \ac{WWL} are activated by connecting to logic `1' for a duration of 2.5 ns.
	This reverses the current direction through the \ac{HM} layer, and eventually resets the \ac{MTJ} along the $\mbox{-}z\mbox{-}$direction.
	The temporal variation of corresponding signal values and the $z\mbox{-}$component of the magnetization, $m_z$, is depicted in Fig~\ref{fig:read_write_reset}(c).
	This indicates that a single \ac{MTJ} completes a full write-read-reset cycle in approximately 6 ns.
	
	\begin{figure*}[!t]
		\centering
		\includegraphics[scale=0.58]{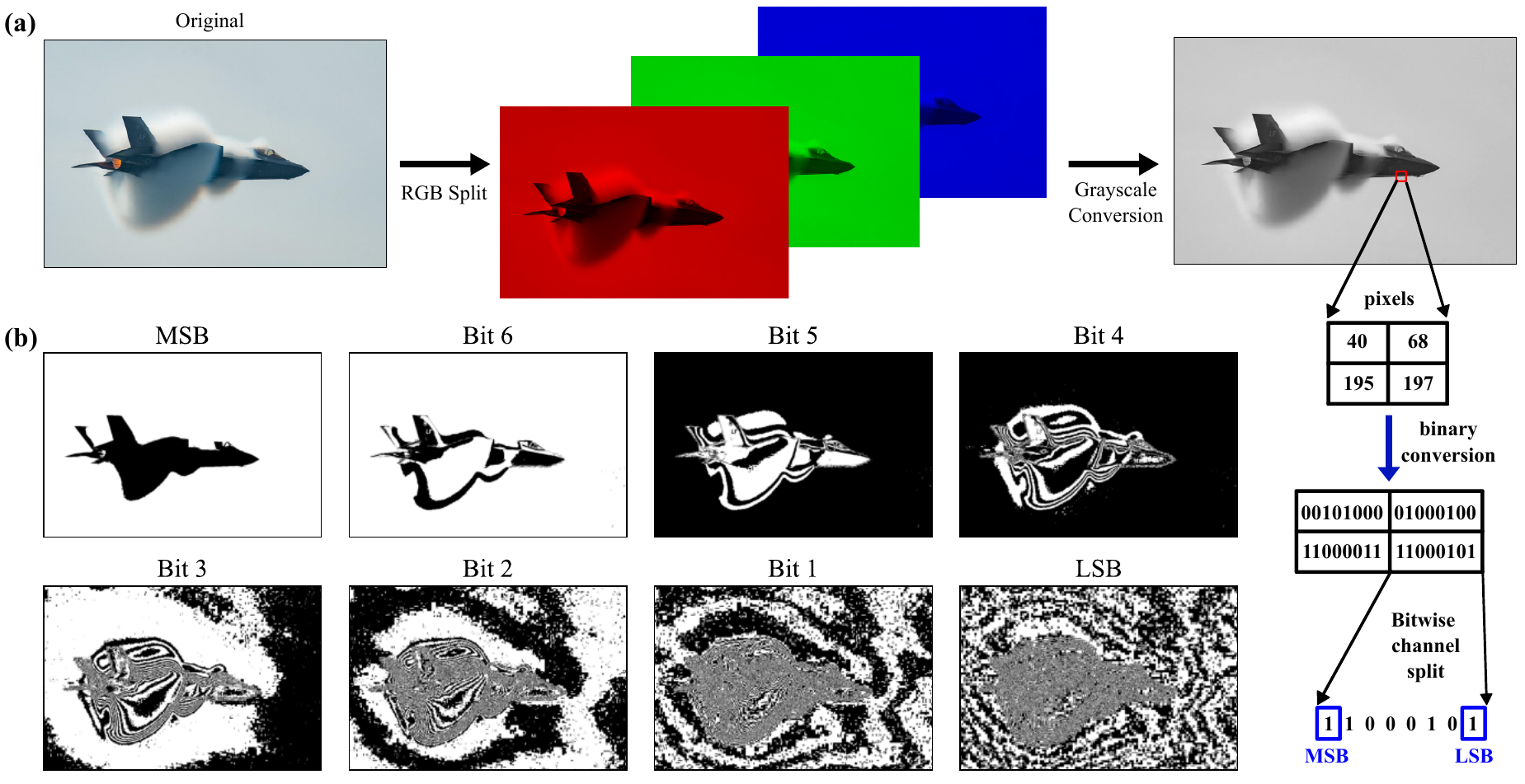}
		\caption{(a) Preprocessing: Original image split to RGB channels, and converted to a single-channel grayscale image. A sample of 2$\times$2 pixels demonstrating binary conversion and bit-wise channel split. (b) Splitting of the grayscale images according to unsigned 8-bit integer channels.}
		\label{fig:RGB}
	\end{figure*}
	
	\begin{table}[!t]
    	\caption{Parameter Values}
    	\label{tab1}
    	\centering
    	\scriptsize
    	\begin{tabular}{@{}lll@{}}
    	\toprule
    	\textbf{Parameter} & \textbf{Symbol} & \textbf{Value} \\
    	\midrule
    	Gyromagnetic Ratio & $\gamma$ & $1.76\times10^{11}\ \mathrm{rad/s\cdot T}$ \\
    	Reduced Planck's Constant & $\hbar$ & $1.054\times10^{-34}\ \mathrm{J\cdot s}$ \\
    	Free Space Permeability & $\mu_0$ & $4\pi\times10^{7}\ \mathrm{T\cdot m/A}$ \\
    	Damping Constant & $\alpha$ & $0.01$ \\
    	Saturation Magnetization & $M_s$ & $650\ \mathrm{kA/m}$ \\
    	Write Current & $I_c$ & $200\ \mu\mathrm{A}$ \\
    	MTJ Dimension (elliptical) &  & $120\times40\times2.5\ \mathrm{nm^3}$ \\
    	HM Layer Dimension &  & $120\times40\times2.8\ \mathrm{nm^3}$ \\
    	Spin-Hall Angle & $\theta_{SH}$ & $0.3$ \\
    	Spin-Flip Length & $\lambda_{sf}$ & $1.4\ \mathrm{nm}$ \\
    	RA Product &  & $10\ \Omega\cdot\mu\mathrm{m}^2$ \\
    	TMR &  & $150\%$ \\
	\bottomrule
	\end{tabular}
	\end{table}

	Next, we calculate the energy consumption corresponding to these three operations.
	Considering the \ac{HM} layer as $\beta\mbox{-}\mathrm{W}$, we assume the resistivity as 100 $\mathrm{\mu \Omega\mbox{-}cm}$, which determines the $\mathrm{R_{HM}}\sim120~\mathrm{\Omega}$.
	We estimate the energy consumption of 12 fJ for both \textit{write} and \textit{reset} operations, as they both take 2.5 ns.
	On the other hand, for the \textit{read} operation, we need to apply a small voltage ($\sim$0.1 V) to sense the current, so that no read disturbance occurs.
	Despite the same read voltage, the read current, $\mathrm{I_P}$ for the P-state, and $\mathrm{I_{AP}}$ for the AP-state would be different, owing to different values of $\mathrm{R_{P}}$ and $\mathrm{R_{AP}}$, as obtained from the RA-product and TMR values given in Table-I.
	A time duration of 1 ns for the \textit{read} operation leads to an energy consumption of 3.7 fJ and 1.5 fJ for the P-state and AP-state, respectively.

	\section{Image Edge detection}

	In this section, we demonstrate the method for image edge detection and compute the corresponding latency and energy consumption.
	Most color images are composed of three channels, Red, Green, and Blue (RGB), where the pixels for each channel are arranged in a two-dimensional grid.
	Edge detection is performed exclusively on single-channel grayscale images; therefore, a preprocessing step is required for any color image.
	Initially, the color image is split into its RGB channels, as shown in Fig.~\ref{fig:RGB}(a).
	To convert it to a single-channel grayscale image, one can either compute the mean of the three channels or use the weighted formula 0.3*R+0.59*G+0.11*B, where R, G, and B represent the respective pixel values.
	We use the latter method in our simulation, as illustrated in Fig.~\ref{fig:RGB}(a).
	The pixel values of grayscale images are 8-bit unsigned integers, which can range from 0 to 255.
	For instance, a sample of 2$\times$2 pixel grid, as shown in Fig.~\ref{fig:RGB}(a), represents these unsigned integer values.
	Subsequently, each pixel's integer value is converted into an 8-bit unsigned binary number, with the \ac{MSB} on the left and the \ac{LSB} on the right.
	This approach can be applied to the entire grayscale image, converting it into eight separate bit channels, ranging from the MSB to the LSB, as depicted in Fig.~\ref{fig:RGB}(b).
	It is noteworthy that the \ac{MSB} contains the maximum amount of image information, which progressively decreases toward the \ac{LSB}.
	Consequently, edge detection can be effectively performed using either the \ac{MSB} alone or a combination of the \ac{MSB} and Bit 6.
	
	\begin{figure}[!b]
		\centering
		\includegraphics[scale=0.55]{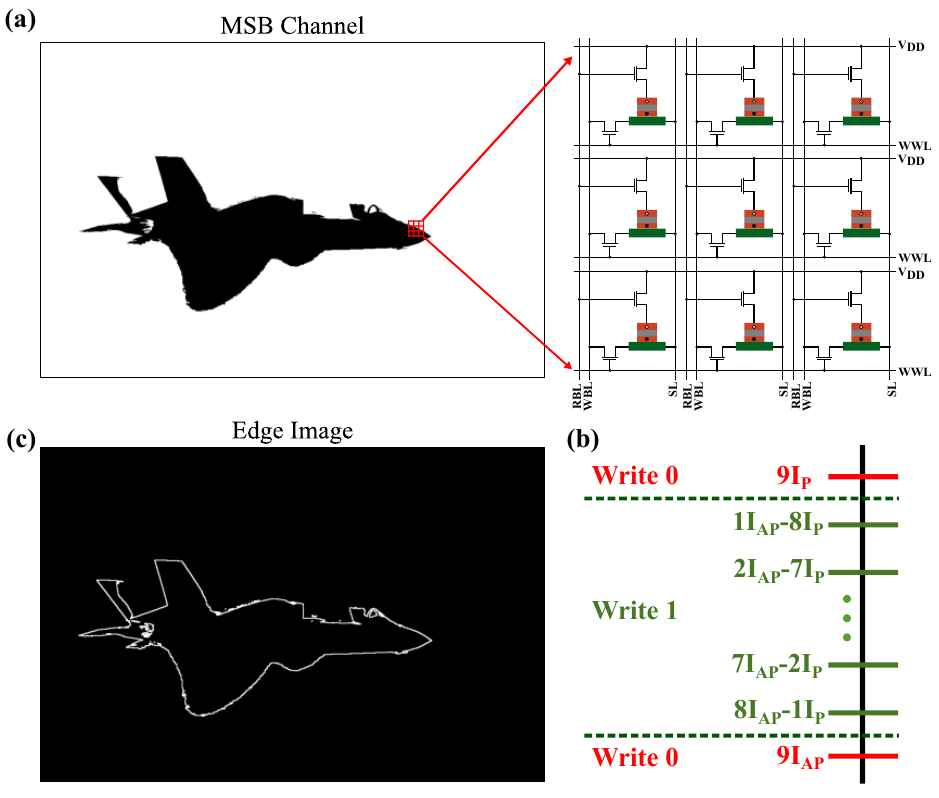}
		\caption{(a) A 3$\times$3 kernel (red squares) is convolved over the MSB channel of the image. Each pixel of the kernel is implemented using a SOT-MTJ bit cell along with two access transistors, shown in enlarged scale.          
        (b) Electrical readout from the kernel, illustrating the three possible current levels—$9\mathrm{I_P}$ for all‐1 input (no edge), $9\mathrm{I_{AP}}$ for all‐0 input (no edge), and intermediate summed current values for mixed inputs (edge detected). 
        (c) The resulting binary edge image produced after thresholding the read current response, highlighting detected object boundaries in the original image.}
		\label{fig:kernel}
	\end{figure}

	For the image shown in Fig.~\ref{fig:RGB}(a), we use only the MSB plane, which is a binary image where the pixel values are either `1' or `0'.
	Next, we implement a 3$\times$3 kernel with the SOT-MTJ bitcells and convolve it over the MSB image with a stride of 1, as shown in Fig.~\ref{fig:kernel}(a)~\cite{lv2024cram}.
	The convolution process involves a three-step cycle-\textit{write}, \textit{read}, and \textit{reset} for each \acp{MTJ} corresponding to each pixel.
	To begin, we apply the pixel values to the \ac{WWL} for 2.5 ns to perform the \textit{write} operation.
	If all nine pixel values are `1', every \ac{MTJ} will switch to the \ac{P}-state.
	Conversely, if all pixel values are `0', none of the \acp{MTJ} will switch and they will remain in the \ac{AP}-state.
	If a subset of the nine pixels has a value of `1', only the corresponding \acp{MTJ} will switch.
	For the subsequent 1 ns, the read operation is performed by simultaneously applying a small voltage across all the \acp{MTJ} while deactivating the \ac{WWL} and \ac{WBL} and activating the \ac{RBL}.
	There are three possible outcomes: (i) If all nine pixel values applied during the write operation were `1', all nine \acp{MTJ} switch to the \ac{P}-state.
	This results in a current of $\mathrm{I_P}$ flowing through each individual \ac{MTJ}, yielding a total current of 9$\mathrm{I_P}$.
	(ii) If all nine pixel values were `0', none of the \acp{MTJ} switch, resulting in a total read current of 9$\mathrm{I_{AP}}$.
	(iii) If a subset of the pixels were `1', the corresponding \acp{MTJ} switch, leading to a read current that falls between 9$\mathrm{I_P}$ and 9$\mathrm{I_{AP}}$, as shown in Fig.~\ref{fig:kernel}(b).
	For the two extreme cases, where all nine pixels are identical, indicating no variation, we conclude no edge is detected.
	Consequently, a pixel value of `0' can be assigned to the edge detection images.
	Conversely, in the third case, where the read current is between 9$\mathrm{I_P}$ and 9$\mathrm{I_{AP}}$, a pixel variation is present, indicating an edge in the binary image.
	Therefore, a pixel value of `1' can be assigned to the edge detection image.
	This process effectively converts the input image into an edge detection image, as depicted in Fig.~\ref{fig:kernel}(c).
	Considering the numerical values computed in Sec. \ref{Sec:Device}, for the \textit{write}, \textit{read}, and \textit{reset} energy for a single SOT-MTJ, the total energy consumption for detecting the edges of the given MSB image (1024$\times$679 pixels) is calculated as 0.16~$\mathrm{\mu J}$, with a latency of 4 ms.
	
	\begin{figure}[!h]
		\centering
		\includegraphics[scale=0.41]{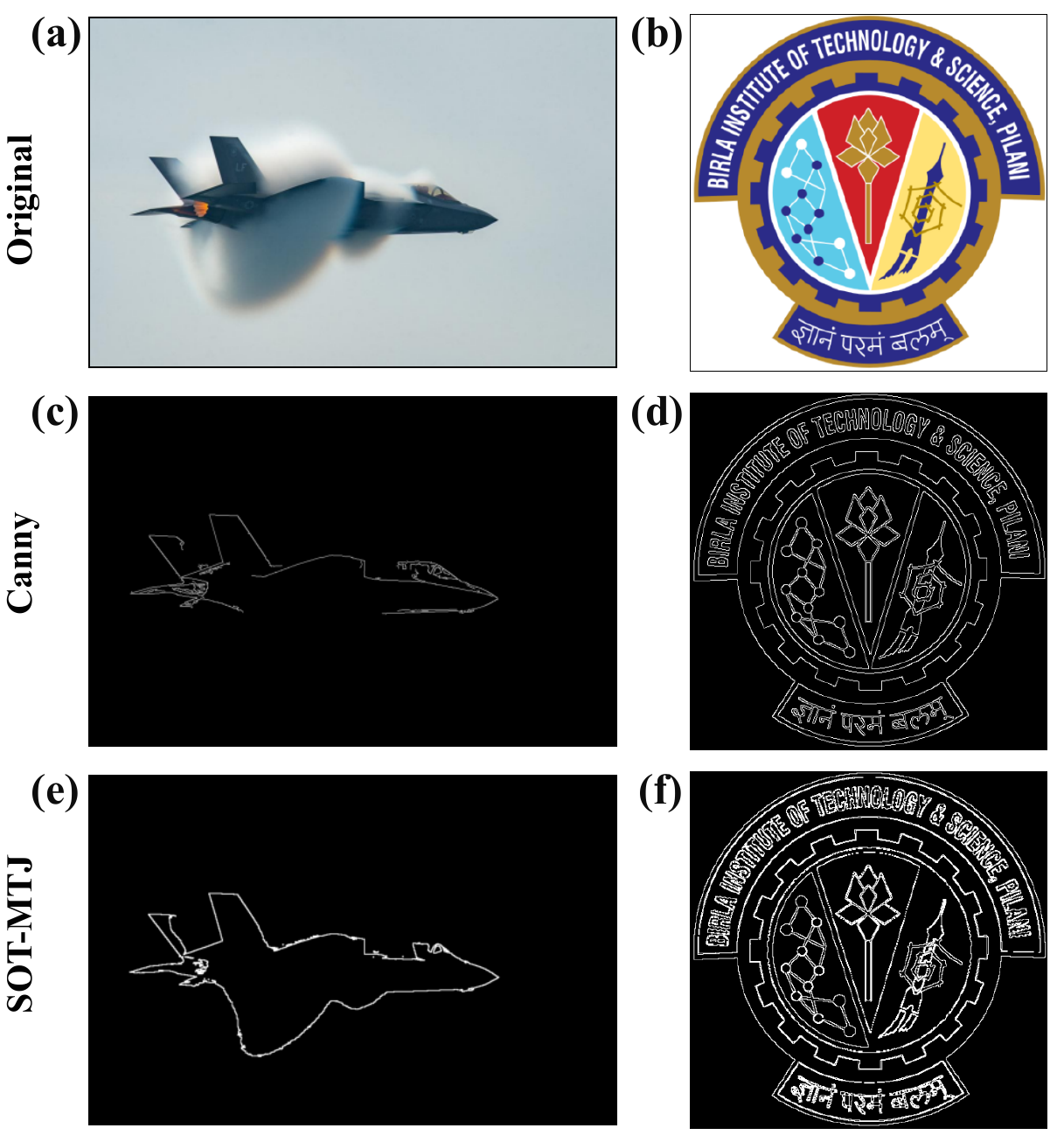}
		\caption{
            Comparison between the proposed SOT-MTJ-based edge detection method and the conventional Canny algorithm.
            (a) and (b) show the original test images used for evaluation. 
            (c) and (d) present the corresponding edge maps generated by Canny operator, (e) and (f) demonstrate the edge detected image using our SOT-MTJ based approach. 
            %
            }
		\label{fig:canny_compare}
	\end{figure}
	
	Next, we compare our approach with the established Canny edge detection algorithm~\cite{canny1986computational}. Fig.~\ref{fig:canny_compare} (a) and (b) displays two distinct images and the rest of the images demonstrate corresponding edge-detected outputs from conventional Canny edge detection algorithm (Fig.~\ref{fig:canny_compare} (c) and (d)) and SOT-MTJ based approach (Fig.~\ref{fig:canny_compare} (e) and (f)). 
    The results show that SOT-MTJ based approach accurately captures high-contrast contours while suppressing minor background details, whereas the Canny output exhibits finer at the cost of computationally intensive method.
	Fig.~\ref{fig:canny_compare}(e) demonstrate SOT-MTJ method successfully identifies a prominent and bright edge, including the air cloud formed due to the sonic boom of the fighter jet, as a distinct object.
	In contrast, the Canny algorithm's output only captures the fighter jet itself and eliminates the white air cloud.
	This difference arises because the Canny algorithm involves a series of complex, sequential preprocessing steps, including Gaussian noise reduction, gradient intensity determination, non-maximum suppression, and hysteresis thresholding, whereas we use an unprocessed raw grayscale image.
	These steps are computationally expensive and power-intensive, making them unsuitable for resource-constrained environments.
	Fig.~\ref{fig:canny_compare}(b) shows the logo of BITS Pilani, a static image with minimal noise.
	Here, the key difference in the results is attributed to the extensive preprocessing used by the Canny algorithm, whereas our approach processes the raw image directly.
	The results demonstrate that our method is effective for edge detection in low-noise images with significantly lower energy consumption and latency.
	Specifically, our approach consumed 0.16$\mu$J and 51 nJ, with corresponding latencies of 4 ms and 1.6 ms, for the images in Fig.~\ref{fig:canny_compare}(a) and (b), respectively.

	\section{Conclusion}
	We have demonstrated a low-energy, hardware-friendly approach for image edge detection.
	Unlike the computationally intensive preprocessing of the Canny algorithm, our method relies on converting color images to grayscale and then splitting them across 8 bits-a process that can be efficiently implemented with an analog-to-digital converter.
	Through numerical simulation, we show that SOT-MTJs can be effectively used as a kernel to detect image edges, consuming only a few $\mu$J of energy and incurring a delay of a few ms.
	With such low latency and energy consumption, our approach is promising for use in a range of applications, including the analysis of complex datasets like medical images.
		
	\bibliographystyle{IEEEtran}
	\nocite{*}
	\bibliography{references} 
	
\end{document}